\documentclass[3p,twocolumn]{elsarticle}

\usepackage{lineno}
\usepackage{color}
\usepackage{graphicx}

\linenumbers

\begin{document}

\title{The MIDAS telescope for microwave detection of ultra-high energy cosmic rays}
\author[usc]{J.~Alvarez-Mu\~{n}iz}
\address[usc]{Universidad de Santiago de Compostela, Departamento de F\'isica de Part\'iculas, \\ Campus Sur, Universidad, E-15782 Santiago de Compostela, Spain}
\author[rio]{E.~Amaral Soares}
\address[rio]{Universidade Federal do Rio de Janeiro, Instituto de F\'isica, \\ Cidade Universitaria, Caixa Postal 68528, 21945-970 Rio de Janeiro, RJ, Brazil}
\author[chi]{A.~Berlin} 
\address[chi]{University of Chicago, Enrico Fermi Institute \& Kavli Institute for Cosmological Physics, \\ 5640 S. Ellis Ave., Chicago, IL 60637, USA}

\author[chi]{M.~Bogdan} 
\author[chi,pra]{M.~Boh\'{a}\v{c}ov\'{a}}
\address[pra]{Institute of Physics of the Academy of Sciences of the Czech Republic, \\
Na Slovance 2, CZ-182 21 Praha 8, Czech Republic}
\author[rio]{C.~Bonifazi}
\author[usc]{W.~R.~Carvalho~Jr.}
\author[rio]{J.~R.~T.~de~Mello~Neto}
\author[chi]{P.~Facal~San~Luis} 
\author[chi]{J.~F.~Genat} 
\author[chi]{N.~Hollon} 
\author[chi]{E.~Mills} 
\author[chi]{M.~Monasor} 
\author[chi]{P.~Privitera} 
\author[rio]{A.~Ramos de Castro}
\author[chi,slo]{L.~C.~Reyes}
\address[slo]{Department of Physics, California Polytechnic State University, San Luis Obispo, CA 93401, USA}
\author[chi]{B.~Rouille~d'Orfeuil}
\author[rio]{E.~M.~Santos}
\author[chi]{S.~Wayne} 
\author[chi]{C.~Williams}
\author[chi]{E.~Zas}
\author[chi]{J.~Zhou}

\date{\today}

\begin{abstract}
We present the design, implementation and data taking performance of 
the MIcrowave Detection of Air Showers (MIDAS) experiment, a large field of view imaging telescope designed to detect microwave radiation from extensive air showers induced by ultra-high energy cosmic rays. This novel technique may bring a tenfold increase in detector duty cycle when compared to the standard fluorescence technique based on detection of ultraviolet photons.   
The MIDAS telescope consists of a 4.5 m diameter dish with a 53-pixel receiver camera, instrumented with feed horns operating in the  commercial extended C-Band (3.4 -- 4.2~GHz).  A self-trigger capability is implemented in the digital electronics. The main objectives of this first prototype of the MIDAS telescope - to validate the telescope design, and to demonstrate a large detector duty cycle  - were successfully accomplished in a dedicated data taking run at the University of Chicago campus prior to installation at the Pierre Auger Observatory.  
\end{abstract}
\maketitle
\linenumbers
\renewcommand{\thefootnote}{\fnsymbol{footnote}}
\section{Introduction}
The origin and composition of Ultra-High Energy Cosmic Rays (UHECRs) remains uncertain~\cite{LetessierSelvon:2011dy}, even after the progress made by the latest generation of experiments~\cite{Abraham:2004dt,AbuZayyad:2000uu}. Due to the strong flux suppression above $10^{19}$~eV~\cite{Abbasi:2007sv, Abraham:2008ru}, very large detection areas are necessary to study cosmic rays at these energies. A future UHECR Observatory based on standard techniques - Surface Detector arrays (SD) and Fluorescence Detectors (FD) -  may be  limited by cost and difficulty of deployment. In this context, radio detection techniques are attractive thanks to the low cost of individual elements, the little maintenance required and  a nearly 100\% detection duty cycle.  

Radio emission in the MHz range from extensive air showers (EAS) has been actively studied in the last decade by the LOPES~\cite{Falcke:2005tc} and CODALEMA~\cite{Ardouin:2005xm} experiments. MHz radio-detection is now well established, and AERA~\cite{Fliescher:2012zz},  in commissioning phase at the Pierre Auger Observatory, will have a sufficiently large instrumented area to explore the potential of this technique for detection of the highest energy cosmic rays.

The use of the microwave (GHz) band for EAS detection was originally pursued by Jelley, Charman and collaborators~\cite{Jelley:1966, Charman:1969} in the late 1960s, but abandoned due to the lack of satisfactory understanding of the emission mechanisms and to limitations in detector technology.  Thereafter it remained mostly unexplored, until recent laboratory measurements~\cite{Gorham:2007af} with particle beams have renewed the interest in this part of the radio spectrum. These measurements suggest that microwave radiation is emitted from the weakly ionized air plasma of free electrons produced by the EAS induced ionization of the atmosphere. The radiation is expected to be continuous and relatively flat in frequency, unpolarized and emitted isotropically, and its intensity to scale with the number of particles of the shower.

\begin{figure}[t]
\centering{
\includegraphics[width=0.9\columnwidth]{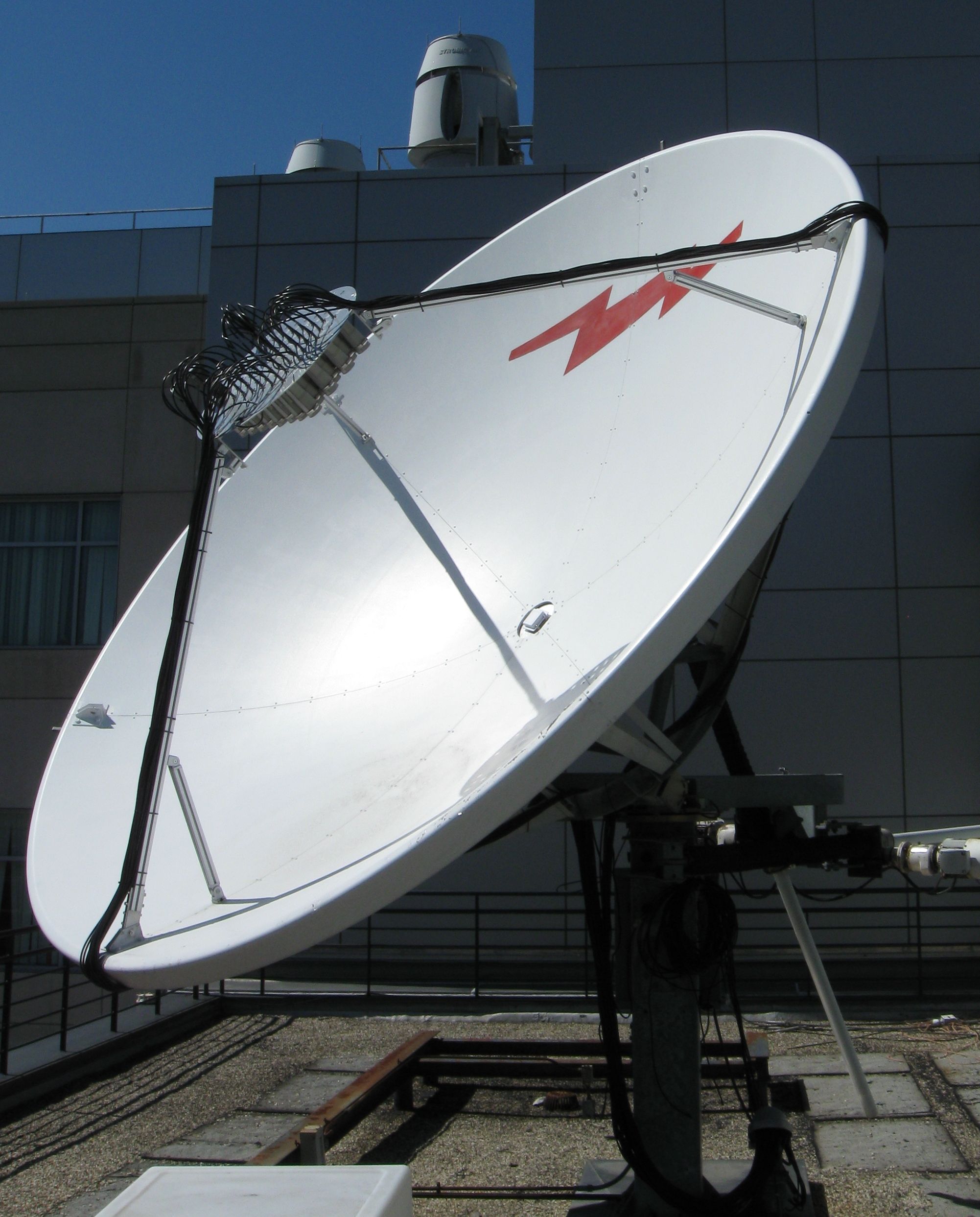} }
\caption{The MIDAS telescope at the University of Chicago, with the 53-pixel camera at the prime
  focus of the 4.5~m diameter parabolic dish reflector.}
\label{fig:telescope}
\end{figure}

Detection of an isotropic emission in the GHz range  - akin to the detection of ultraviolet fluorescence photons by FD pioneered by the Fly's Eye experiment~\cite{Baltrusaitis:1985mx}  and currently used by the Pierre Auger Observatory~\cite{Abraham:2009pm} and the Telescope Array~\cite{Tokuno:2012mi} - allows for the measurement of the EAS development in the atmosphere, which provides a calorimetric measurement of the energy and crucial information on the mass composition of primary cosmic rays.
 A GHz radio telescope would overcome the limitations of the FD technique, i.e. data taking only during moonless nights ($\approx15\%$ duty cycle) and significant systematic uncertainties introduced by light attenuation in the atmosphere. In fact, microwave detectors can operate 100\% of the time, and attenuation in the GHz range is minimal, even with rain or clouds.  
 Moreover, commercial off-the-shelf GHz equipment, mostly developed for satellite TV reception, is readily available and inexpensive. 
 Microwave telescopes could provide existing UHECR experiments of unprecedented sensitivity to primary comic ray  composition, and be employed in a future large scale observatory.  
 
Several complementary approaches to microwave detection of EAS are currently being pursued, including the AMBER and EASIER detectors~\cite{Allison:2011zz} at the Pierre Auger Observatory, and the CROME experiment at KASCADE~\cite{Smida:2011cv}. Also, new laboratory measurements with particle beams are being performed~\cite{MAYBE,AMY} to better characterize the microwave emission.  

In this paper, we present the  MIcrowave Detection of Air Showers (MIDAS) experiment, an imaging telescope whose primary objective is to  confirm the microwave emission from EAS, and to demonstrate the feasibility of a low cost design for this novel technique.  
The MIDAS telescope is described in  Sec.~2. The electromagnetic simulations of the telescope response are illustrated in Sec.~3. The calibration procedures and the measured sensitivity of the instrument are presented in Sec.~4.  A realistic simulation of the MIDAS detection of EAS is described in Sec.~5. Operation and data taking performance of the MIDAS telescope are presented in Sec.~6, and conclusions are drawn in Sec.~7.

\begin{figure}[t]
\centering{
\includegraphics[width=0.9\columnwidth]{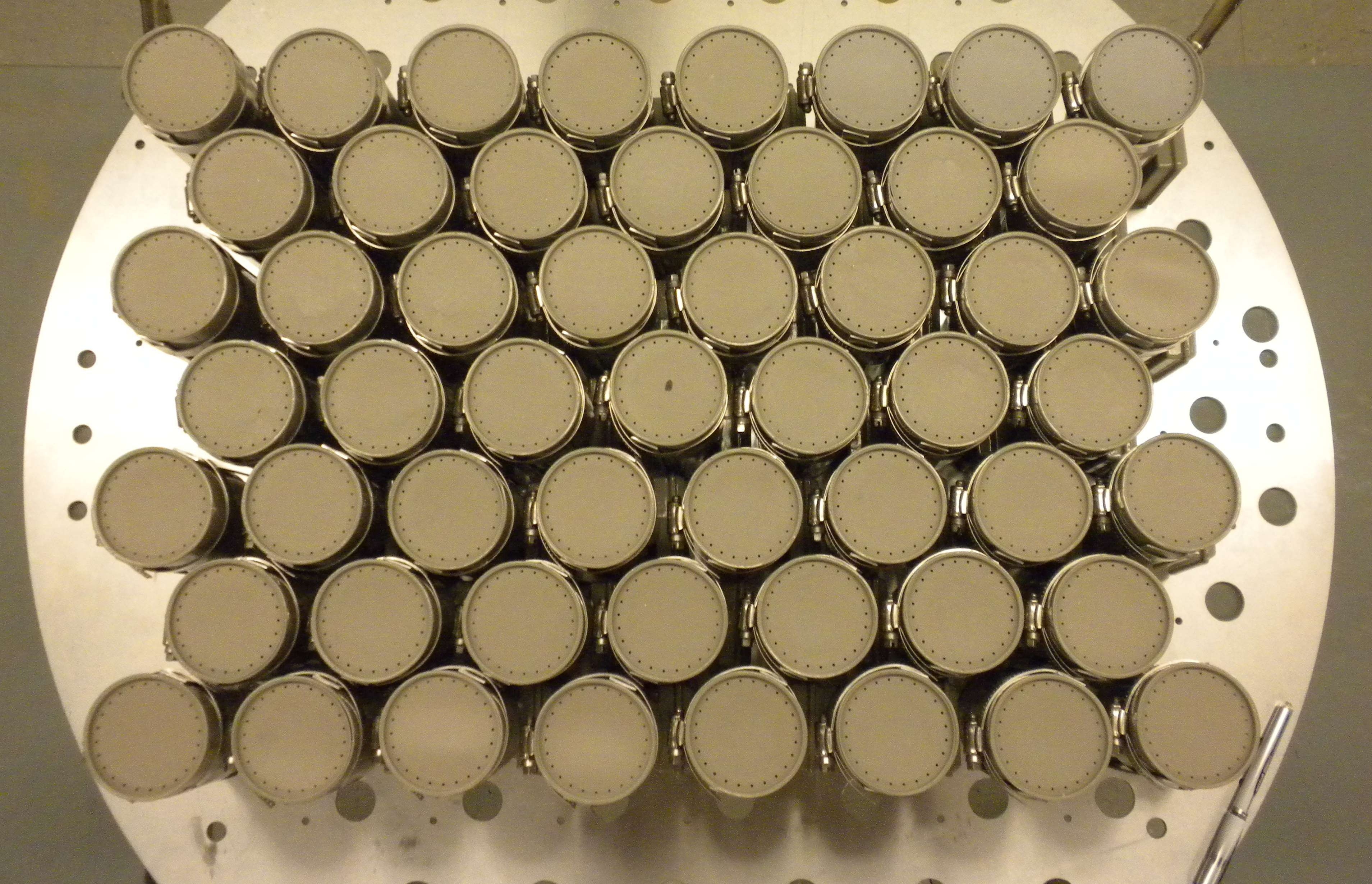}}
\caption{A front view of the receiver camera, with LNBFs closely packed to maximize the sensitivity over the focal plane. }
\label{fig:camera}
\end{figure}

\section{The MIDAS telescope}
The telescope consists of a large parabolic dish reflector with a receiver camera at its prime focus, installed on the roof of the Kersten Physics Teaching Center at the University of Chicago (Fig.~\ref{fig:telescope}). In the following, the design and technical implementation of the different components of the MIDAS telescope are described. 

\subsection{Reflector and receiver camera}
The parabolic dish reflector (Andrew) has  4.5 m diameter and  $f/D=0.34$. A motorized alt-azimuth mount allows for telescope movements in a range of 90$^\circ$ in elevation and $~120^\circ$ in azimuth. The remote control of the telescope pointing ($0.1^\circ$ precision) is  integrated in the data acquisition (DAQ) system (Sec.~\ref{sec:daq}). 
A 53-pixel receiver camera is mounted on the prime focus of the dish, covering a field of view of about $20^\circ \times 10^\circ$. The microwave receivers are arranged in seven rows,
and staggered  to maximize the sensitivity across the focal plane (Fig.~\ref{fig:camera}). 

\begin{figure}[!t]
\centering{
\includegraphics[width=\columnwidth]{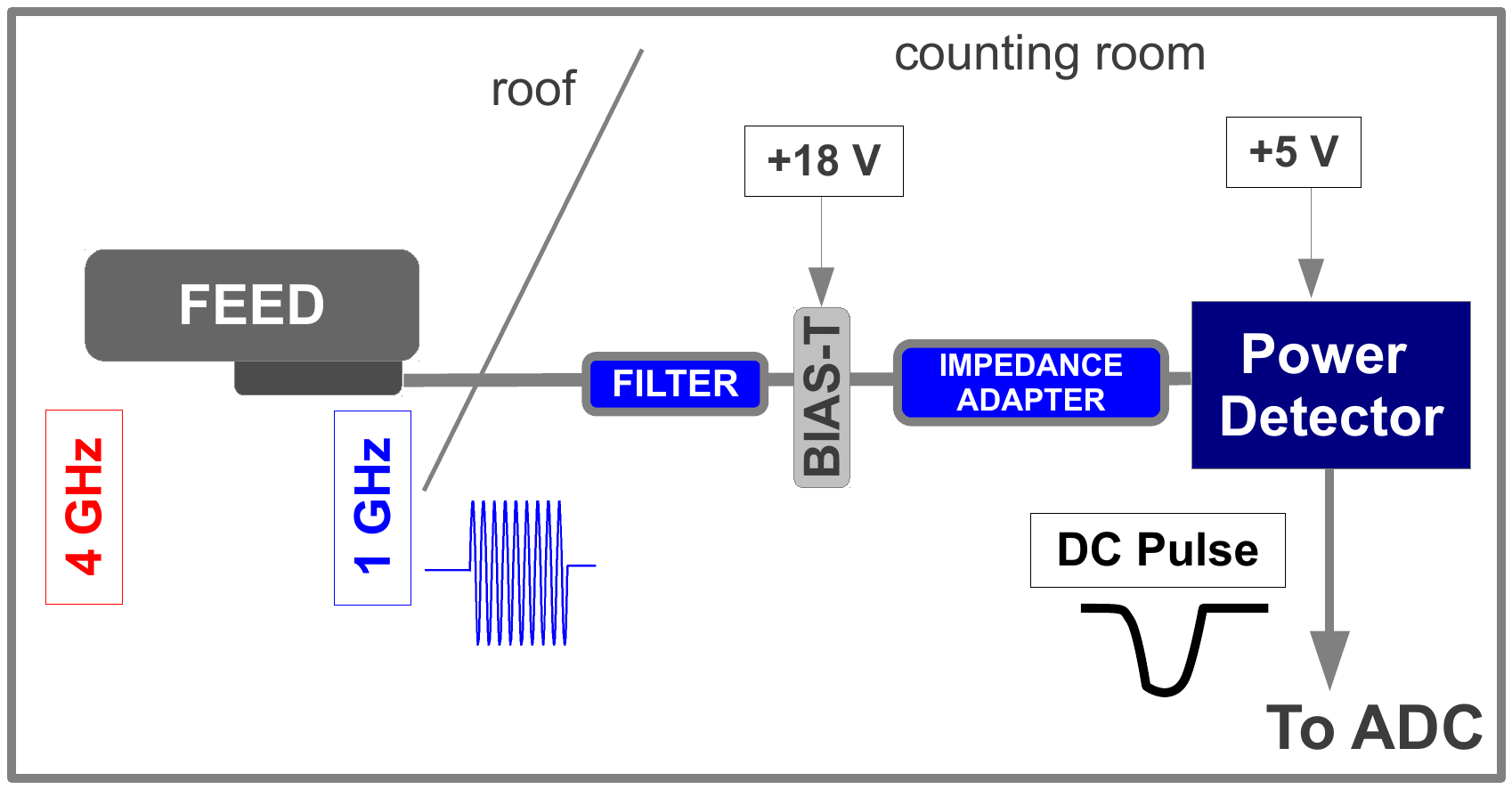}}
\caption{The analog electronics chain. See text for details on the different components.}
\label{fig:analog_channel}
\end{figure}

A commercial low noise block feed horn (LNBF) operating in the extended
C-band (3.4 - 4.2 GHz) is used for the receiver. These feeds (WS International) are mass-produced for
consumer satellite television.  The LNBF integrates  a feed horn, low noise
amplifiers, and a frequency downconverter.  The feed can receive two
orthogonal linear polarizations which are remotely selectable through the LNBF power voltage level setting. 
A 5150 MHz local oscillator in the frequency downconverter mixes the input RF 
signal down to a frequency interval of  950-1750 MHz, which is 
transmitted with minimal loss through standard  
coaxial cable. The receiver bandwidth, its gain $\Gamma$ and noise temperature were measured to be about 1 GHz,  65~dB and 20~K, respectively.

Power to the LNBF is provided through 30 meters of commercial quad-shielded RG-6 coaxial cable, which also brings the RF signal, after amplification and downconversion, from the telescope to the counting room.  
 
 \subsection{Analog electronics}
 \label{subsec:analog}
 The analog electronics chain is summarized in Fig.~\ref{fig:analog_channel}. The RF signal in the coaxial cable is first passed through a bandpass filter (1.05-1.75 GHz).
The purpose of the filter is to reject interference from radar altimeters of airplanes, which was identified as a major source of background in the early stage of commissioning of the telescope.  A power inserter (commonly called bias tee) provides the DC voltage to the LNBF, while letting through the RF signal. An impedance adapter matches the standard 75 Ohm impedance of these commercial satellite TV  components to the 50 Ohm impedance of the RF power detector (Mini-Circuits  ZX47-60-S+). The overall power loss $L$ in the 30 m of cable and in the analog electronics was measured to be 17 dB.   

\begin{figure}[!t]
\centering{
\includegraphics[width=0.9\columnwidth]{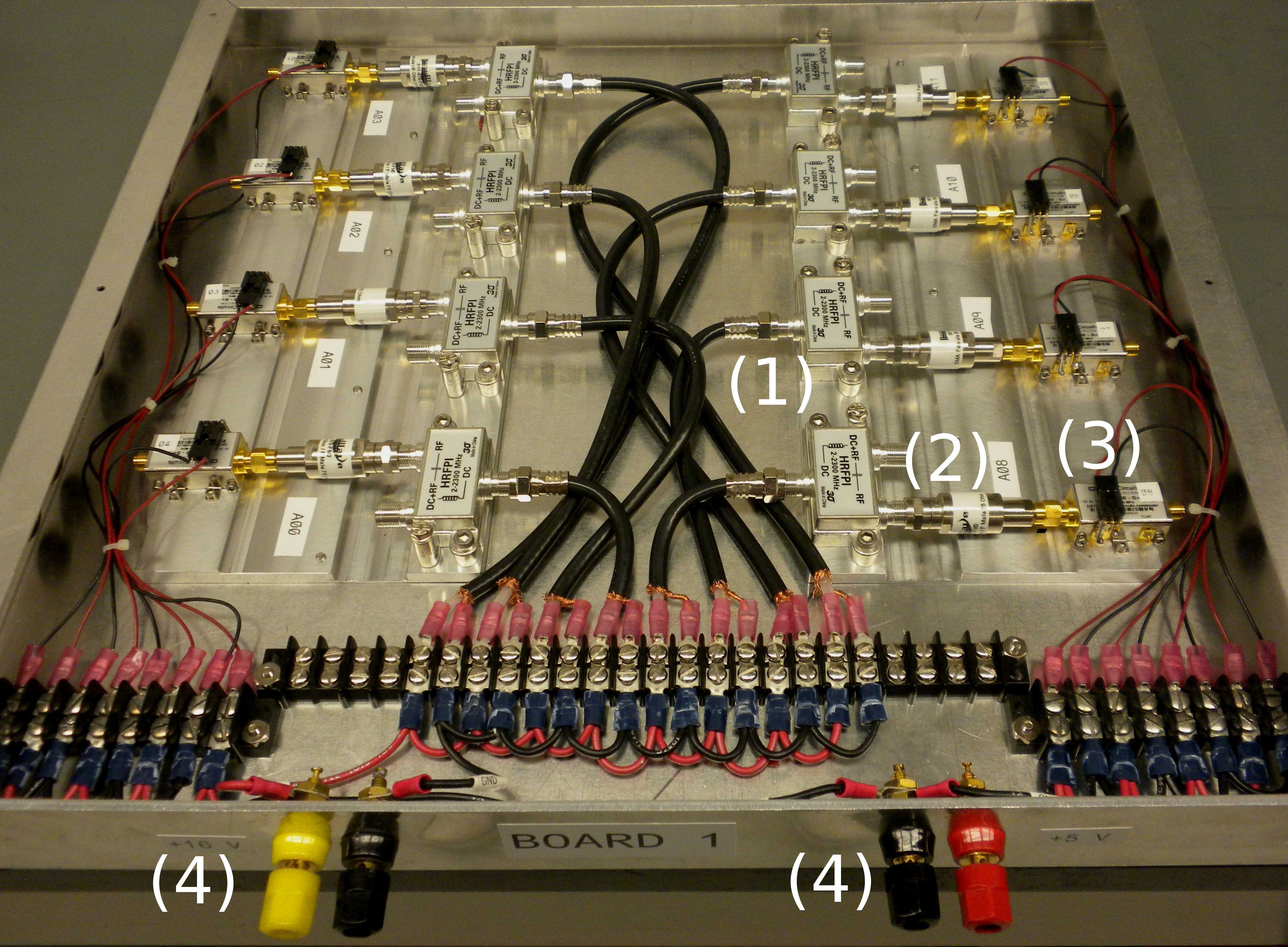}}
\caption{An analog electronics tray with eight channels. The bias tee (1), impedance adapter (2), power detector (3), and DC voltage inputs (4) are visible. }
\label{fig:analog_tray}
\end{figure}

The power detector (input bandwidth $\approx 8$~GHz)  responds 
logarithmically to an RF power $P$ in the range -55 dBm to 5 dBm, with a voltage output between 2.0~V and 0.5 V: 
\begin{equation}
V = V_0 - 10\, a  \log(P),
\label{eq:pwdetector}
\end{equation} 
with $P$ in mW. The characteristics of the 53 power detectors were individually measured, with typical values around  $V_0=0.625$~V and $a=0.025$~V/dB. 
Also, their response time was found to be about 100~ns, well suited for typical pulses of microseconds duration expected from an EAS crossing the field of view of a pixel. 

The analog electronics components are arranged in trays - 8 channels per tray - for  distribution of  DC power and routing of the signals to the digital electronics  (Fig.~\ref{fig:analog_tray}). 

\begin{figure}[ht]
\centering{
\includegraphics[width=0.9\columnwidth]{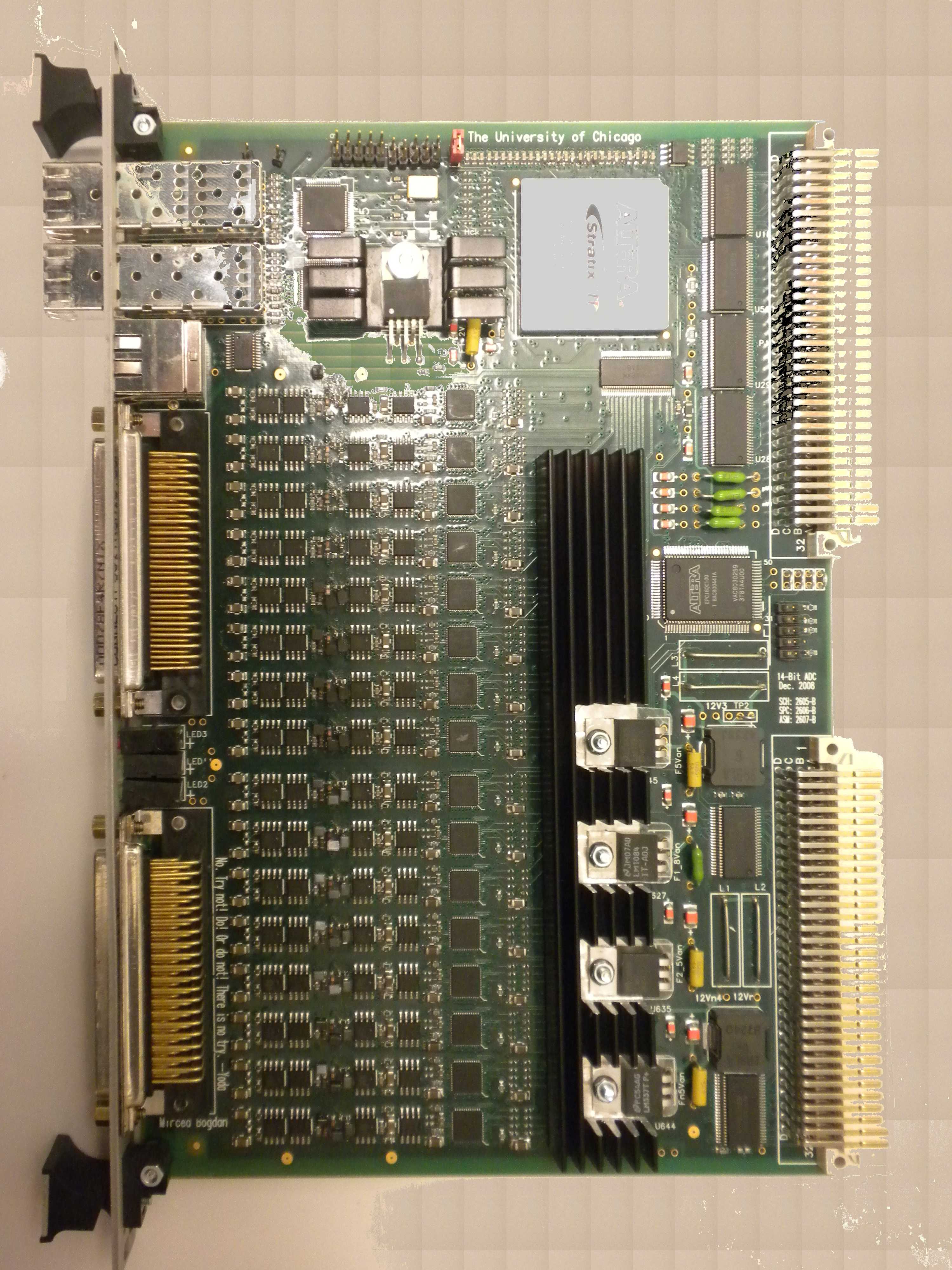}}
\caption{Digital electronics board with 16 Flash-ADC channels. The module
  includes an FPGA for the First Level Trigger logic and a VME interface. The board 
  was designed by the Electronics Design Group at the Enrico
  Fermi Institute of the University of Chicago.}
\label{fig:fadc}
\end{figure}

\subsection{Front-End electronics and digitizers}
\label{subsec:fadc}
The signal from the power detector is  digitized with 14-bit resolution (calibration constant ${b=7}$~ADC/mV) at a sampling rate of 20 MHz 
by custom-made Flash-ADC (FADC) boards (Fig.~\ref{fig:fadc}). Up to 2048 samples 
are stored in a circular buffer 
and processed by a first level trigger algorithm implemented in the on-board  FPGA. A standard VME interface allows for board control and data readout. 

Four FADC boards, each with 16 channels, are sufficient for the entire receiver camera.   
A Master Trigger Board, also equipped with FPGA logic and VME interface, provides the global clock for the FADC synchronization and  performs a high level trigger decision.
Each FADC board is connected to the Master Trigger board through Low Voltage Differential
Signaling (LVDS) lines carrying the clock and trigger signals.

A VME module equipped with a GPS receiver (Hytec 2092) tags the
time of the triggered events with 10~ns precision.
Details of the MIDAS digital electronics can be found in \cite{Mircea}.

\begin{figure}[!t]
\centering{
\includegraphics[width=\columnwidth]{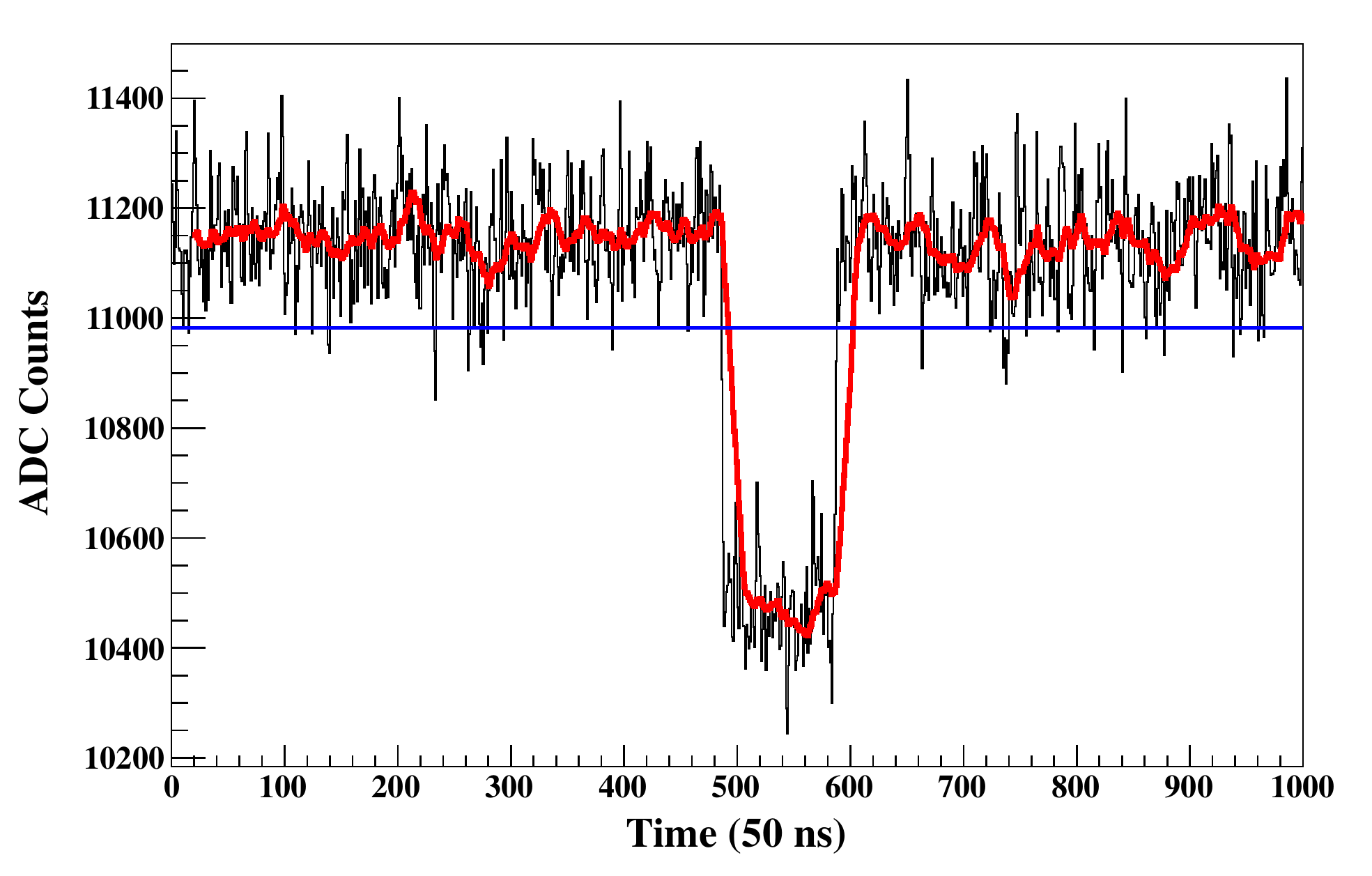}}
\caption{ Illustration of the FLT. The digitized time trace for  a 5 $\mu$s RF pulse from the calibration antenna (see Sec.~\ref{sec:daq}), with the ADC running average of 20 consecutive time samples superimposed as a gray histogram (red in the color version). An FLT is issued when the running average falls below the threshold, indicated by the horizontal line. }
\label{fig:flt}
\end{figure}

\begin{figure}[!th]
\centering{
\includegraphics[width=0.3\textwidth]{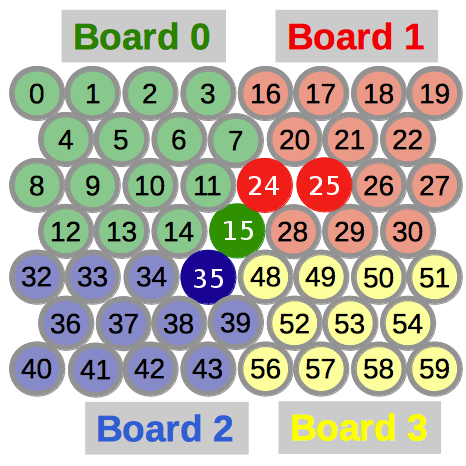}
\includegraphics[width=0.4\textwidth]{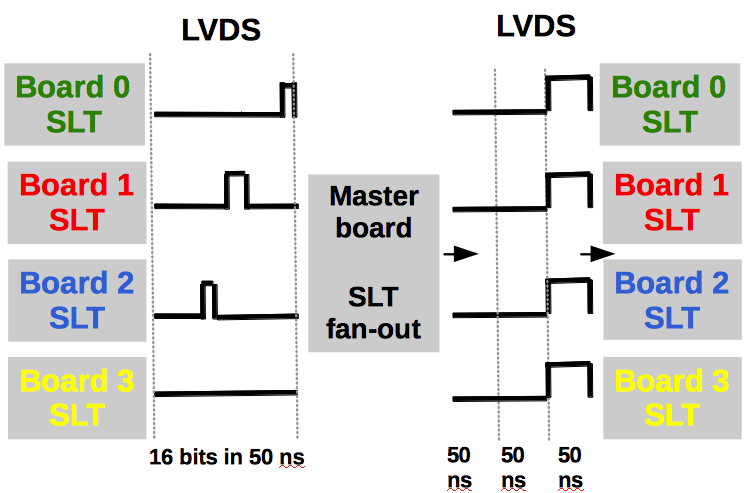}}
\caption{Illustration of the SLT. The 4 highlighted pixels (top panel), located in three different FADC boards, have an FLT occurring in the same 50 ns time sample. Each FADC board sends through the LVDS lines the FLT status of its pixels to the Master Trigger Board (bottom panel, left), where a matching SLT pattern is found. An SLT is then sent back through the LVDS lines to the FADC boards (bottom panel, right) triggering the entire camera for data readout. }
\label{fig:slt}
\end{figure}

\begin{figure*}[!tbh]
\centering{
\includegraphics[width=\textwidth]{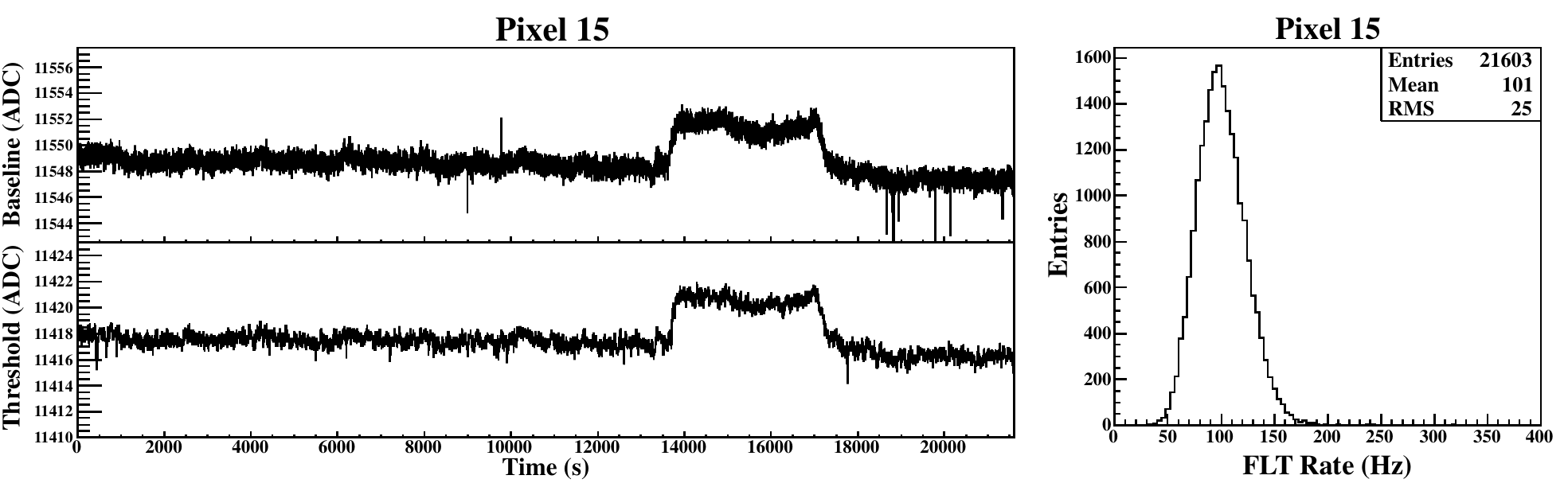}}
\caption{Monitoring data from one pixel for a six-hour data taking run. On the left, the ADC baseline averaged over 10~ms (upper panel) and the FLT threshold (bottom panel) as a function of time. The threshold follows closely the variations of the baseline, keeping the FLT rate during the run (right panel) around 100~Hz. }
\label{fig:monitoring}
\end{figure*}

\subsection{Trigger}
\label{sec:trigger}
MIDAS implements a multi-level trigger architecture, optimized for transient events with topology and time structure compatible with signals from an EAS. The system is modeled on the one successfully operating at the FD of the Pierre Auger Observatory~\cite{Abraham:2009pm}. For each channel, a First Level Trigger (FLT) algorithm identifies pulses in the FADC trace. The FLT information from all channels is sent  to the Master Trigger Board where the Second Level Trigger (SLT) searches for patterns of FLTs compatible with a cosmic ray track.

The FLT status bit of a given channel is activated whenever the ADC  running sum of 20 consecutive time samples falls below a threshold (Fig.~\ref{fig:flt}). To compensate for changes in the noise level, the threshold is continuously regulated in order to keep a stable FLT rate around 100 Hz. The active status of the FLT bit is extended for $10~\mu$s after the trigger, to allow for time coincidences between different channels in the Master Trigger Board.  
 
Every 50 ns, each FADC board transmits 16 bits, corresponding to the FLT status of its 16 channels, to the Master Trigger Board (Fig.~\ref{fig:slt}), where  
the SLT trigger algorithm searches for 4-fold
patterns of channels with FLT triggers overlapping in time. There are
767 patterns compatible with the topology of a cosmic ray shower (straight tracks across the camera).
When an SLT matching pattern is found, an event  trigger signal is
distributed back to the FADC boards and to the GPS module. At this point, 
 a block  of 100 $\mu$s of ADC data (including 500 pre-trigger
samples) is frozen in the memory buffer and
made available for readout via VME. All 53 channels, even those not participating to the trigger decision,  are readout. 

Higher levels of trigger control have also been implemented. For example, the FLT threshold
regulation and data acquisition are inhibited whenever the SLT rate is higher
than a certain limit, to be automatically restored when the SLT rate comes back to a normal level. 

\subsection{Data acquisition, monitoring and operation}
\label{sec:daq}
A single board computer (GE Intelligent Platforms V7865) acting as VME master is used for the data acquisition. The DAQ software monitors the
relevant VME registers of the FADC boards, and reads out through the VME bus the GPS time stamp and the data available in the buffer after an event trigger.
The event data are then assembled and written to disk in ROOT format~\cite{Brun:1997pa}.

For monitoring purposes, the ADC baseline averaged over 10 ms (calculated in the FPGA of the FADC board), the FLT rate, and the FLT threshold of each channel are readout and recorded every second. An example of monitoring data collected during a six-hour data taking run is presented in Fig.~\ref{fig:monitoring}.

Additional monitoring information is obtained from RF pulses periodically illuminating the receiver camera. For this purpose, a patch antenna with a wide beam (HPBW=$70^\circ$) is mounted at the center of the reflector dish. The antenna is driven by an RF signal generator located in the counting room, which every 15 minutes produces a train of ten RF pulses of  10~$\mu$s width. These data provide monitoring information for all of the channels at the same time, ensuring
they are properly operating and that the trigger is working as designed.

\begin{figure*}[th]
\centering{
\includegraphics[width=0.4\textwidth]{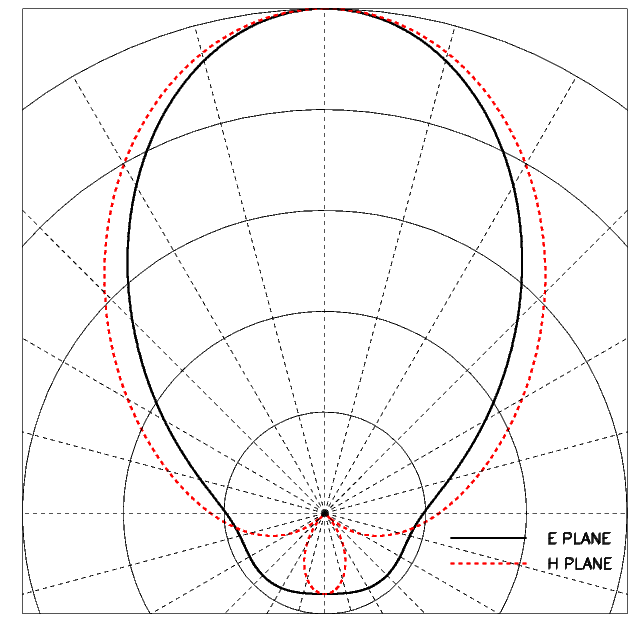}
\includegraphics[width=0.4\textwidth]{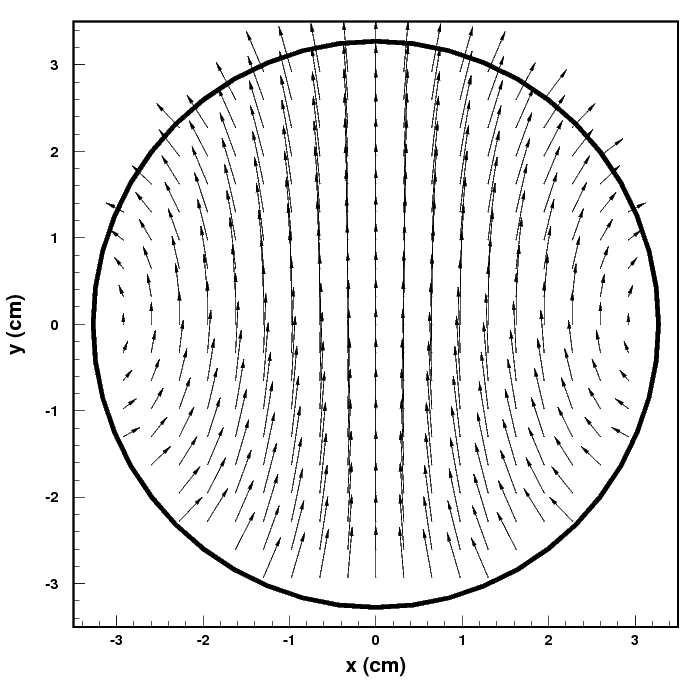}}
\caption{EM calculation of the feed for the dominant TE$_{11}$ mode at 3.8 GHz. Left: Normalized electric field radiation pattern for the MIDAS feed in polar coordinates. 
Both the E-plane ($\phi=\pi/2$) and H-plane are shown ($\phi=0$). Right: Electric field vector over the feed aperture.}
\label{fig:feedPP}
\end{figure*}

Data runs last six hours and are automatically restarted, with data files  backed up on a server for offline analysis. A fully automatic and remote operation of the telescope was achieved through a slow control software for the antenna positioning, and for the status of power supplies and of the VME crate.

\section{EM simulation of telescope efficiency}
\label{sec:optics_sim}
The efficiency of the MIDAS telescope has a significant angular dependence, due to the large field of view implemented in a parabolic reflector. For a proper calibration of the telescope and for a realistic estimate of its sensitivity to EAS, the power pattern of each pixel - i.e the pixel detection efficiency as a function of the direction $(\theta, \phi)$ of the incident microwave radiation with respect to the telescope boresight -  must be determined. A full electromagnetic (EM) simulation of the reflector and receiver camera has been developed for this purpose. 

At radio and microwave wavelengths, it is convenient to simulate the antenna system (reflector and feed) in emission mode, making use of  the so-called Reciprocity Theorem \cite{kraus, orfanidis} to obtain its effective area $A_i$ in a given direction: 
\begin{equation}
\label{recip}
A_i(\theta,\phi)=\frac{\lambda^2}{4\pi}G_i(\theta,\phi),
\end{equation}
where $\lambda$ is the wavelength of the radiation and $G_i(\theta,\phi)$ is the gain of the antenna system for pixel $i$, defined as the power emitted per unit solid angle in the direction $(\theta, \phi)$ normalized to the corresponding power emitted by an isotropic radiator. 
In practical terms, the antenna effective area summarizes the dependence of the telescope gain on the dimensions of the reflector, the frequency of the radiation, and overall illumination (i.e. taper) of the feed. 

\begin{figure*}[!t]
\centering{
\includegraphics[width=\textwidth]{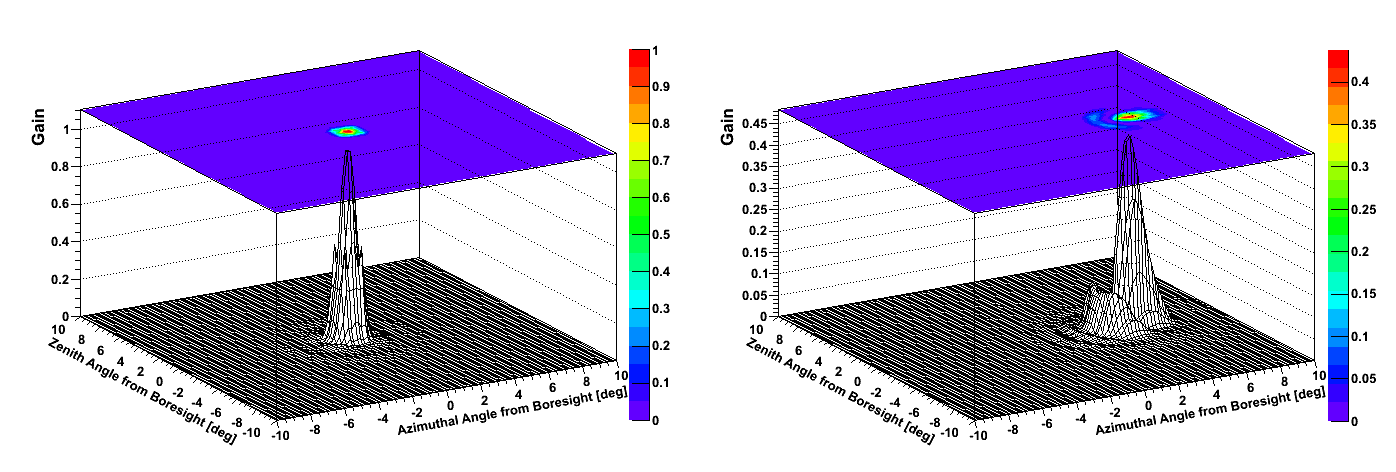}}
\caption{Left: Power radiation pattern for the central feed (pixel 15) of the MIDAS camera. This feed presents the maximum antenna gain of the receiver camera, and a symmetrical pattern with very small sidelobes. Right: Radiation pattern for a feed laterally displaced by 16.5 cm ($\sim2\lambda$) from the camera focus (pixel 13). Its gain  is $\sim$50\% smaller, and presents a significant coma lobe.}
\label{fig:feedsGain}
\end{figure*}

The feed is simulated as two cylindrical waveguides -  the one corresponding to the feed aperture having a diameter ${d=6.7}$~cm and the other having a diameter ${d=5.5}$~cm - joined by a small conical throat.  The dominant propagating mode is the TE$_{11}$ (electric field perpendicular to the feed axis) with a cutoff wavelength of 9.4~cm determined by the diameter of the end section ($\lambda_{TE_{11}}^c=1.71d$) \cite{silver}. The calculated radiation pattern of the feed in polar coordinates for the E-plane ($\phi=\pi/2$) and H-plane ($\phi=0$), and the direction of the electric vector field over the feed aperture are shown in Fig.~\ref{fig:feedPP}. 

To obtain the antenna effective area, the electromagnetic field distribution over the telescope aperture is calculated. For each feed, the telescope aperture is found by intersection of rays emitted by the feed with the plane perpendicular to the telescope axis and containing the dish focal point.
The paths of rays from the feed to the aperture plane are determined in the optical limit using Fermat's principle of least transit time. Spherical wave propagation is assumed from feed to dish and plane wave propagation is assumed from dish to aperture. A proper treatment of the optical paths is essential for MIDAS, since the lateral displacement of the feeds with respect to the dish focal point produces significant differences in their gain. Also, the shadow of the receiver camera, which increases the size of sidelobes, is included in the simulation.
The radiation field is then calculated in the far-field limit by the Fourier transform of the aperture field distribution, and the power is obtained as the square of the associated Poynting vector.

Equation \ref{recip} holds for an antenna system completely matched to its transmission line.
In practice, reflections at the feed entrance, at the throat between the cylindrical sections and at the coupling between the feed and the coaxial line reduce the antenna effective area. Their effect can be calculated introducing an equivalent transmission line model with impedance mismatches which produce reflected power waves. Additional signal losses in the coaxial cable and in the analog electronics are included in the simulation. Also, the reflector does not collect all the power emitted by the feed, and part of it spills over, further reducing the antenna total effective area.

The result of the EM simulation can be expressed as a relative power pattern for each pixel $i$: 
\begin{equation}
\label{epsilon}
  \epsilon_i(\theta,\phi) = \frac{G_i(\theta,\phi)}{G_{15}(0,0)},
\end{equation}
where $G_{15}(0,0)$ is the calculated gain of the central pixel in the direction of boresight. The pattern for the central feed (Fig.~\ref{fig:feedsGain}, left) is symmetric around the boresight with relatively small sidelobes. Feeds away from the focal point (Fig.~\ref{fig:feedsGain}, right) have smaller gains and bigger, asymmetrical lobes. From Eq.~\ref{recip}, the effective area of the antenna system to a microwave flux incident from a given direction is then given by: 
\begin{equation}
\label{recip1}
A_i(\theta,\phi)=A_{eff} \, \epsilon_i(\theta,\phi),
\end{equation}
where $A_{eff} = \frac{\lambda^2}{4\pi}G_{15}(0,0)$ is the effective area of the dish. 

\begin{figure}[!tbh]
\centering{
\includegraphics[width=0.4\textwidth]{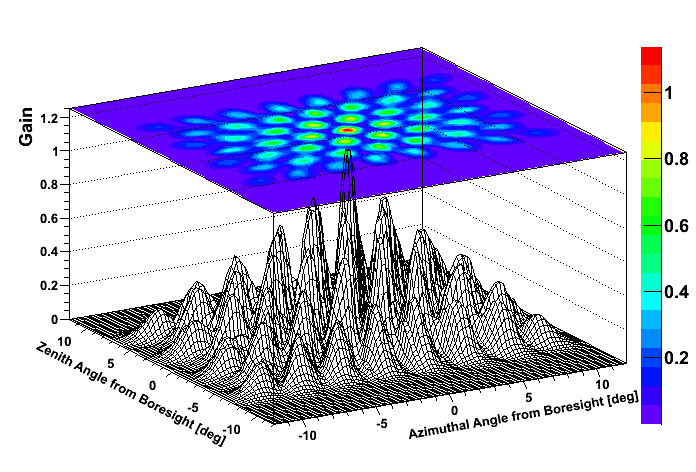}}
\caption{Radiation pattern of the MIDAS camera, calculated by summing over all pixels, as function of the angle with respect to the antenna boresight.}
\label{fig:CameraGain}
\end{figure}

The overall sensitivity of the camera can be calculated summing over the pixels: $\epsilon(\theta,\phi) = \sum_{i} \epsilon_{i}(\theta,\phi)$, and is presented in Fig.~\ref{fig:CameraGain}. Even with the compact arrangement of the MIDAS feeds over the focal plane, there is a region of reduced efficiency between adjacent feeds.
Also, the efficiency at the borders of the field of view is as low as 20\% of the efficiency at the center. 
While not uniform over the camera, the telescope efficiency is reasonably high in a large portion of its field of view, which is sufficient for the goals of this first low-cost prototype for the MIDAS concept.

\section{Telescope Calibration}
\label{sec:absolute_calibration}
Dedicated calibration measurements were performed during the commissioning of the
telescope, which were followed up during data taking to verify the stability of the system.

The Sun, whose flux in the microwave range is measured daily by several solar radio observatories around the world, was used as the primary source for the absolute calibration of the MIDAS telescope. This calibration was cross-checked with other known astrophysical sources. 

The monitoring patch antenna (Sec.~\ref{sec:daq}) was also used for relative channel calibration.

\subsection{Absolute calibration}
An unpolarized microwave flux density, $F$, incident on the MIDAS reflector from a direction $(\theta,\phi)$ produces, at the input of the power detector of a given pixel $i$, a power $P_i$ given by: 
\begin{eqnarray}
P_i &=&\frac{ L_i  \cdot \Gamma_i \cdot B_i \cdot A_{eff}\cdot \epsilon_i(\theta,\phi)\cdot F }{2}\nonumber \\ 
& = & \alpha_i  \cdot \epsilon_i(\theta,\phi) \cdot F,
\end{eqnarray}
where $\epsilon_i(\theta,\phi)$ is the pixel relative power pattern, $A_{eff}$ is the effective area of the dish, $B_i$ is the bandwidth of the electronics (essentially given by the pixel bandpass filter), $\Gamma_i$ is the gain of the receiver, $L_i$ is the loss due to the 30 m cable and the analog electronics, and the factor 2 takes into account the fact that the linearly polarized receivers detect only half of the available power.    
The output voltage of the power detector (Eq.~\ref{eq:pwdetector}) is digitally sampled in the FADC board, yielding a signal $n_i$ in ADC counts:
\begin{equation}
  n_i = n^0_i - 10\, k_i \log\left[ \alpha_i \cdot \epsilon_i(\theta,\phi) \cdot  F\right],
\label{eq:f}
\end{equation}
where $n^0_i = b V_0$, and $k_i=a \, b \simeq 175$ ADC/dB with $a$, $V_0$  and $b$ as defined in Sec.~\ref{subsec:analog} and \ref{subsec:fadc}, respectively. The calibration constant $k_i$ is independently measured for each channel (see Sec.~\ref{sec:relcal}).

 Even when the telescope is pointed towards a region of the sky with no microwave sources, a power $P_i^{sys}$ is present at the input of the power detector, with a corresponding signal $n_i^{sys}$ in ADC counts. 
 Sources for this signal include radiation from the sky in the main beam of the antenna, radiation from ground in the antenna sidelobes, and  electronics noise, dominated by the receiver noise temperature. An equivalent flux density,  $F_i^{sys}$, is defined from $P_i^{sys}=  \alpha_i F_i^{sys}$.
 
A measured signal $n_i$ can be converted into an absolutely calibrated flux density $F$ by deriving from Eq.~\ref{eq:f}:
\begin{equation}
n_i  = n_i^{sys} -  10\, k_i \log \left(1+\frac{\epsilon_i(\theta,\phi) \cdot F}{F_i^{sys}}\right)\!\!,
\label{eq:Fcal}
\end{equation}
 where $\epsilon_i(\theta,\phi)$ is taken from the simulation described in Sec.~\ref{sec:optics_sim}, and  $n_i^{sys}$ and the corresponding $F_i^{sys}$ must be determined through calibration procedures. 

\begin{figure*}[t]
\centering{
\includegraphics[width=\textwidth]{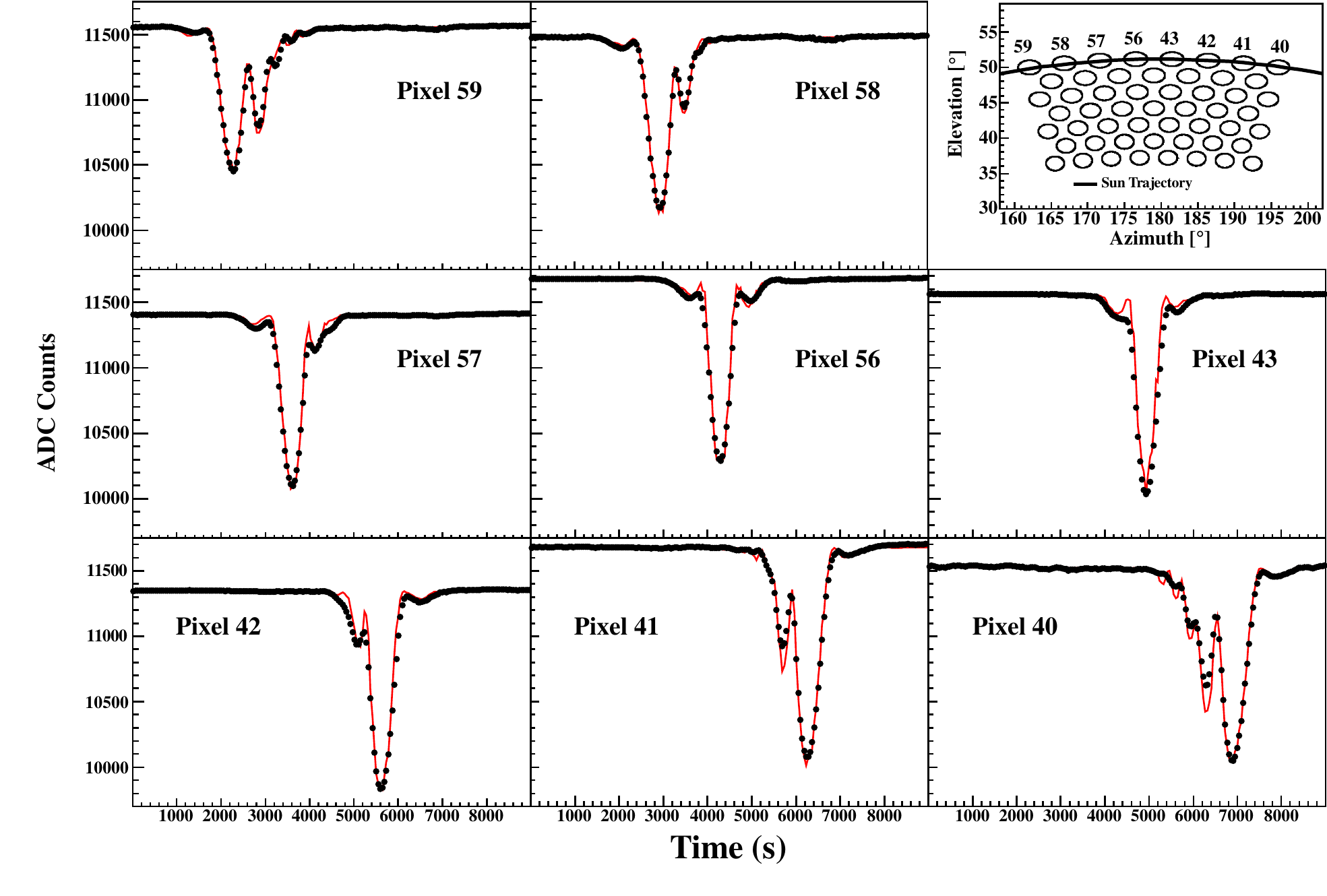}}
\caption{Example of a calibration measurement of the MIDAS telescope with a Sun transit. Top right inset: Sun trajectory in the top row of the camera pixels (azimuth measured clockwise from North). The measured ADC signal (closed dots) for each pixel is shown in the other panels, with the superimposed line representing the EM simulation for the best value of $F_i^{sys}$ (see text). }
\label{fig:SunTransit_vs_Expectation}
\end{figure*}

The Sun was used as the main calibration source of the MIDAS telescope, by measuring the pixel signal during a transit of the Sun in its field of view. From Eq.~\ref{eq:Fcal}, the time evolution of the signal is given by: 
\begin{eqnarray}
\lefteqn{n_i(t)   = n_i^{sys} -}  \nonumber \\ 
& &  10\, k_i \log \left(1+\frac{\epsilon_i(\theta(t),\phi(t)) \cdot F_{sun}}{F_i^{sys}}\right)\!\!,
\label{eq:Fcal1}
\end{eqnarray}
where $n_i^{sys}$ is the ADC baseline measured before the Sun transit, $(\theta(t),\phi(t))$ is the sun position in the sky at time $t$, and the Sun flux density, $F_{Sun}$, is taken from the daily measurements by the Nobeyama observatory~\cite{nobeyama}. The value of $F_i^{sys}$ which best describe the calibration data is then obtained.
   An extensive measurement campaign was performed, with each Sun transit aiming to the calibration of a row of pixels. An example of a Sun calibration run for the top row of pixels is presented in Fig.~\ref{fig:SunTransit_vs_Expectation}. These measurements not only provide an absolute calibration of each pixel through $F_i^{sys}$, but also demonstrate the quality of the simulation of $\epsilon_i(\theta,\phi)$.
 In fact, simulations were found to describe very well the pixel power pattern in both the width and the relative efficiency of the lobes, even for pixels at the edges of the camera where strong aberrations are present. 

As an example, several calibration measurements of the central pixel yielded a $F_{15}^{sys}=1.96\cdot 10^4$~Jy\footnote{1 Jy = $10^{-26}$ W/m$^2$/Hz}, equivalent, in the common notation of radio astronomy, to a system temperature $T_{sys}$ of:   
\begin{equation}
T_{sys}= \frac{F_{15}^{sys}\cdot A_{eff}}{2 k_b} = \rm{65~K},
\label{eq:tsys}
\end{equation}
where $k_b$ is the Boltzmann constant, and $A_{eff} = 9.1$~m$^2$  is the effective area ($\approx 60\%$ of the geometric area) of the telescope as determined by simulations described in Sec.~\ref{sec:optics_sim}. 
The other pixels had a similar system temperature. 
The estimated systematic uncertainty on the measurement of $F_{i}^{sys}$ is 15\%, dominated by the uncertainty of the Nobeyama measurement and of the telescope pointing. 

 Measurements of the Moon ($F_{Moon}\simeq F_{Sun}/100$) and of the Crab Nebula ($F_{Crab}\simeq F_{Sun}/1000$) were also performed, providing a cross-check of the absolute calibration. Using the Sun calibration, 
the Moon temperature was found to be 245~K, to be compared with 268~K~\cite{moon} (the measurement was taken a day after the full moon). 
The flux density of the Crab Nebula was measured to be  $760$~Jy, to be compared with $718 \pm 43$~Jy at 3.38~GHz~\cite{Baars:1977zz}. Both of these results agree to better than 10\%, within the estimated systematic uncertainty of the Sun calibration. 

\begin{figure}[!t]
\centering{
\includegraphics[width=0.9\columnwidth]{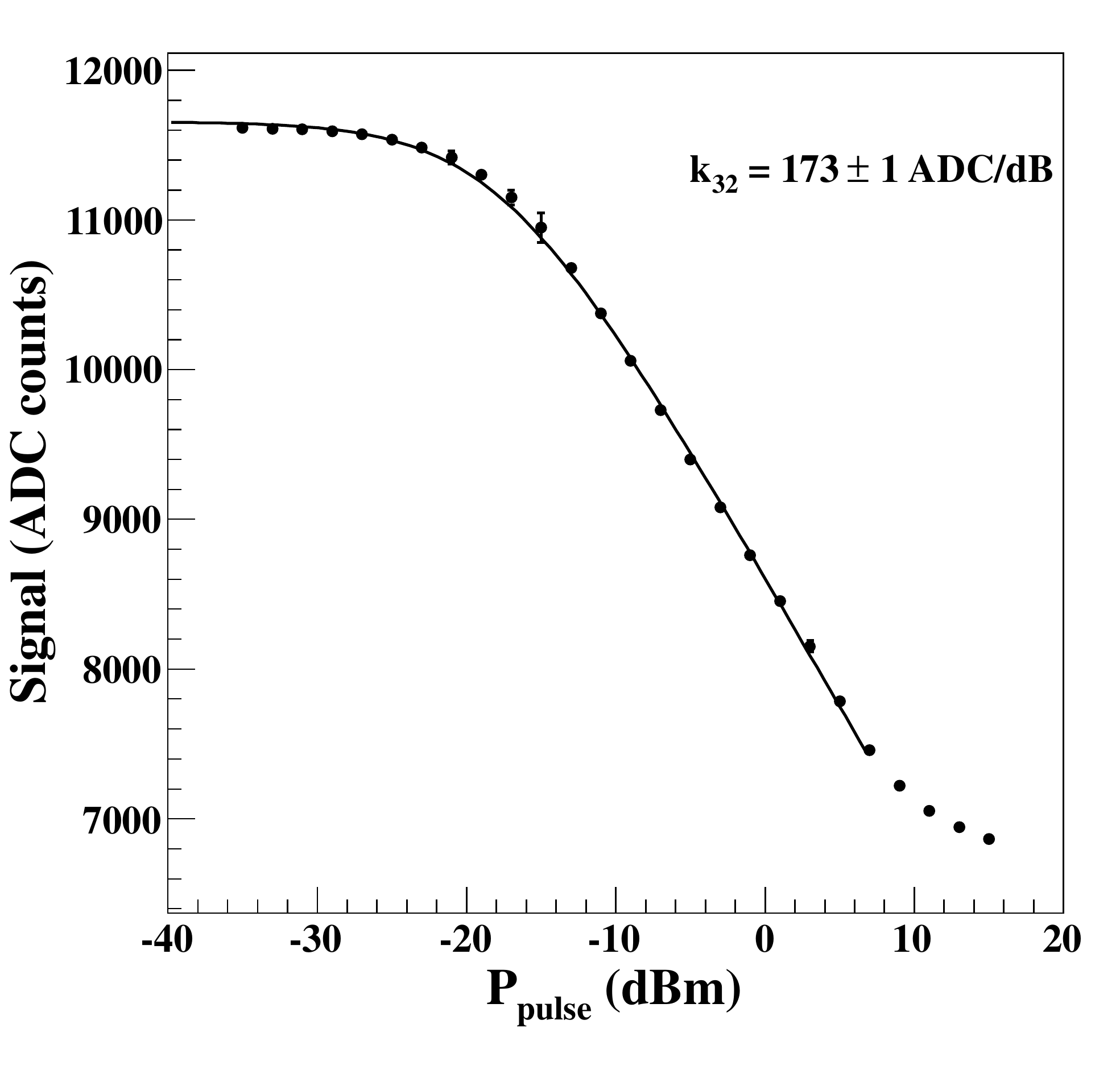}}
\caption{Example of a measurement of the calibration constant $k$. 
Closed dots represent the measured signal in pixel n.~32 for different powers of the calibration antenna pulses. The line is the result of a fit to derive $k_{32}$. For  \protect$P_{pulse}$ greater than  8~dBm, measurements are affected by  saturation of the LNBFs and were excluded from the fit.}
\label{fig:calfit}
\end{figure}

\begin{figure}[!t]
\centering{
\includegraphics[width=0.9\columnwidth]{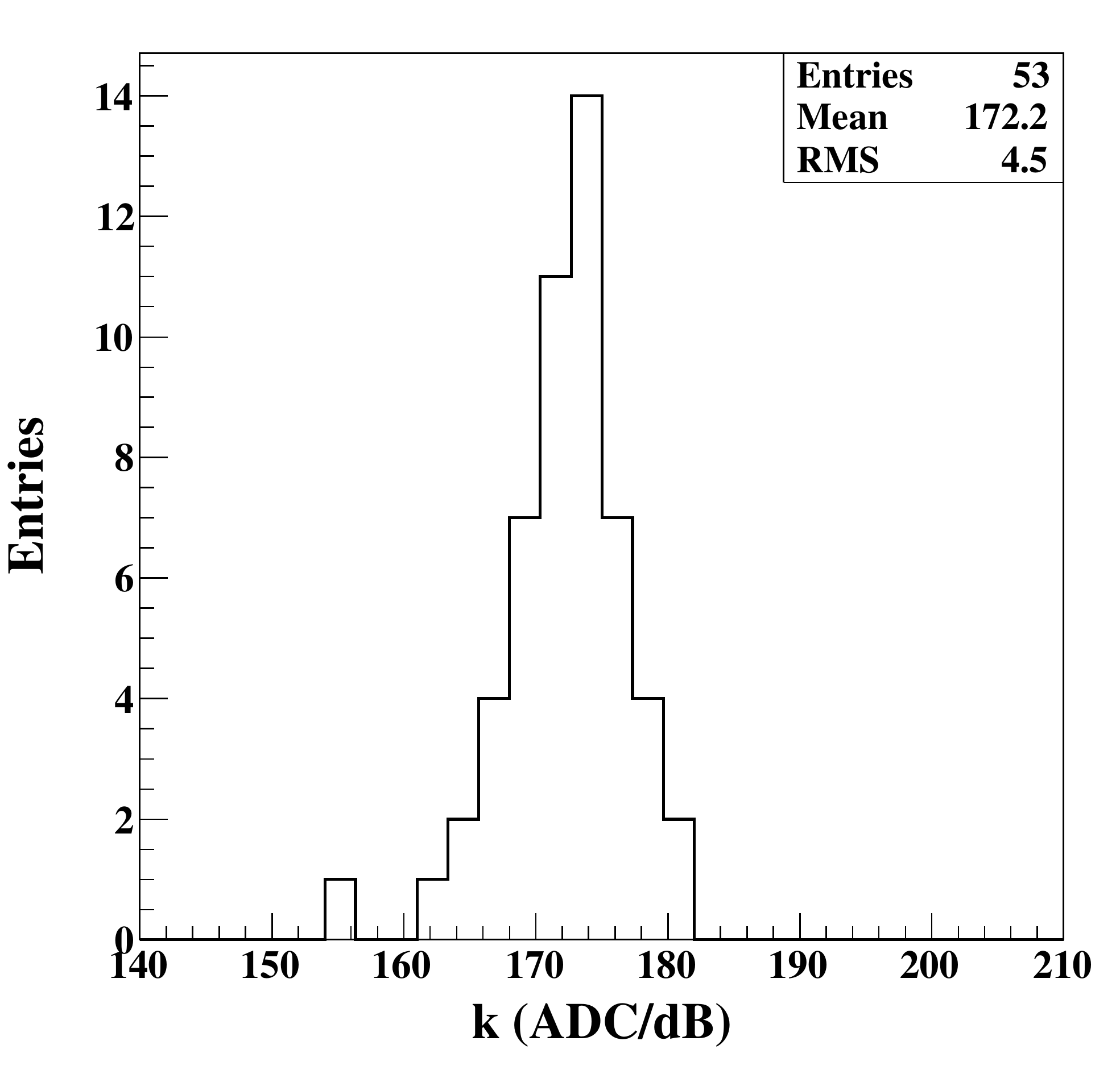}}
\caption{Distribution of the measured calibration constants $k$.}
\label{fig:calconst}
\end{figure}

\subsection{Pixel calibration constants and timing}
\label{sec:relcal}
The calibration constant $k_i$ in Eq.~\ref{eq:f} was measured  by using the patch antenna mounted at the center of the dish. The pulse power $P_{pulse}$ of the RF signal generator driving the patch antenna was changed over a wide range, and the corresponding signal $  n_i^{pulse}$ was measured for each channel.  From  Eq.~\ref{eq:Fcal}, one expects:
\begin{equation}
  n_i^{pulse} = n_i^{sys} - 10\, k_i\log\left(1+\frac{f_i \cdot P_{pulse}}{P_i^{sys}}\right)\!\!,
\label{eq:calrel}
\end{equation}
where $n_i^{sys}$ is the signal measured when the RF signal generator is off, $P_i^{sys}$ is the corresponding power at the input of the power detector, and $f_i \cdot P_{pulse}$ is the power induced by the calibration pulse at the input of the power detector.
The factor $f_i$ may change from pixel to pixel, depending on the distance of the pixel from the patch antenna, on the relative orientation of the linear polarizations of the emitter antenna and the receiver, and on the signal loss $L$.  An example of a fit of  Eq.~\ref{eq:calrel} , with $k_i$, $n_i^{sys}$ and $f_i/P_i^{sys}$ as free parameters, to the calibration data for one channel is shown in  Fig.~\ref{fig:calfit}. The spread of the distribution of the fitted calibration constants for the 53 channels (Fig.~\ref{fig:calconst}) is 2.6\%, with an average of $k=172$~ADC/dB. 

The calibration pulses have a fast risetime (\textless 1~ns) and illuminate all the camera receivers simultaneously, allowing for a measurement of their time response and synchronization. The time at which the amplitude of the measured pulse reaches 50\% of its maximum value, $t_{50}$, was taken as an estimator. The distribution of  $t_{50}$ was found to have an RMS value of 25~ns, more than adequate for the typical pulses of microseconds duration expected from an EAS crossing the field of view of a pixel.

\section{Sensitivity to EAS}
\label{sec:shower_sim}
The sensitivity of the MIDAS telescope to microwave emission from EAS has been studied with Monte Carlo simulations. The shower development in the atmosphere - i.e. the number of charged particles $N$ at any given atmospheric depth - is simulated with a Gaisser--Hillas parameterization \cite{GH}, including proper fluctuations of the shower profile. The shower arrival direction is isotropically distributed, with its landing point randomly distributed over a large area around the telescope. For a given shower geometry, the number of particles, $N(t)$, along the shower profile is calculated, where $t$ is the time in 50~ns samples of the MIDAS digital electronics.

The microwave flux density at the detector aperture is then calculated, following the model in~\cite{Gorham:2007af}, as:
\begin{equation}
 F(t) =  F_{\rm{ref}} \cdot \frac{\rho(t)}{\rho_{0}} \cdot \left(\frac{d}{R(t)}\right)^{2} \cdot \left(\frac{N(t)}{N_{\rm{ref}}}\right)^{\alpha}, 
\label{eq:emission}
\end{equation}
where $F_{\rm{ref}}$ is the flux density at a distance $d=0.5$~m from a reference shower of $E_{ref} = 3.36\times10^{17}$~eV, $R(t)$ is the distance between the detector and the EAS segment, and $\rho (t)~(\rho_{0})$ is the atmospheric density at the altitude of the EAS segment (at sea level). $N_{\rm{ref}}$ is the average number of shower particles at the maximum of the EAS development for a proton primary of energy $E_{ref}$.   The parameter $\alpha$ characterizes phenomenologically the coherence scaling relationship for the EAS microwave emission, with  $\alpha=2$ ($\alpha=1$) corresponding to a fully coherent (incoherent) emission.  Laboratory measurements~\cite{Gorham:2007af} suggest a reference flux density  $F_{\rm{f,ref}}^0=1.85\times10^{-15}\ \rm{W/m^2/Hz}$ and full coherence at shower maximum.  

\begin{figure}[t]
\centering{
\includegraphics[width=\columnwidth]{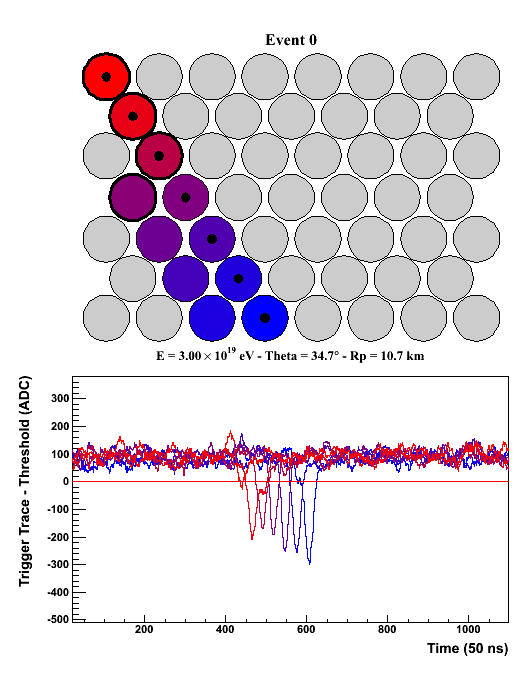}}
\caption{Event display of a $3\cdot 10^{19}$~eV simulated shower landing approximately 10~km from the telescope. In the top panel, pixels with an FLT are highlighted, with color coded by arrival time. In the bottom panel, ADC  running averages of 20 consecutive time samples for the selected pixels (identified by black dots in the top panel) are shown. The running average of each pixel is referred to the threshold level (horizontal line) for display purposes.}
\label{sim_event}
\end{figure}

In order to convert the microwave flux density at the detector aperture into a signal in ADC counts, the efficiency maps  and calibration constants described in Sec.~\ref{sec:optics_sim} and Sec.~\ref{sec:absolute_calibration}  were implemented in the simulation. For each channel, the actual value of $n_{sys}$ and its fluctuation were taken to be equal to their average values measured during several months of data taking, providing a realistic simulation of the telescope sensitivity. The FLT and SLT algorithms of Sec.~\ref{sec:trigger} were also implemented, and all simulated events fulfilling the SLT condition are written to disk in the same format as the data. An example of event simulated with $F_{\rm{ref}} = F_{\rm{ref}}^0$ and $\alpha=1$ is shown in Fig.~\ref{sim_event}.

\begin{figure}[t]
\centering{
\includegraphics[width=\columnwidth]{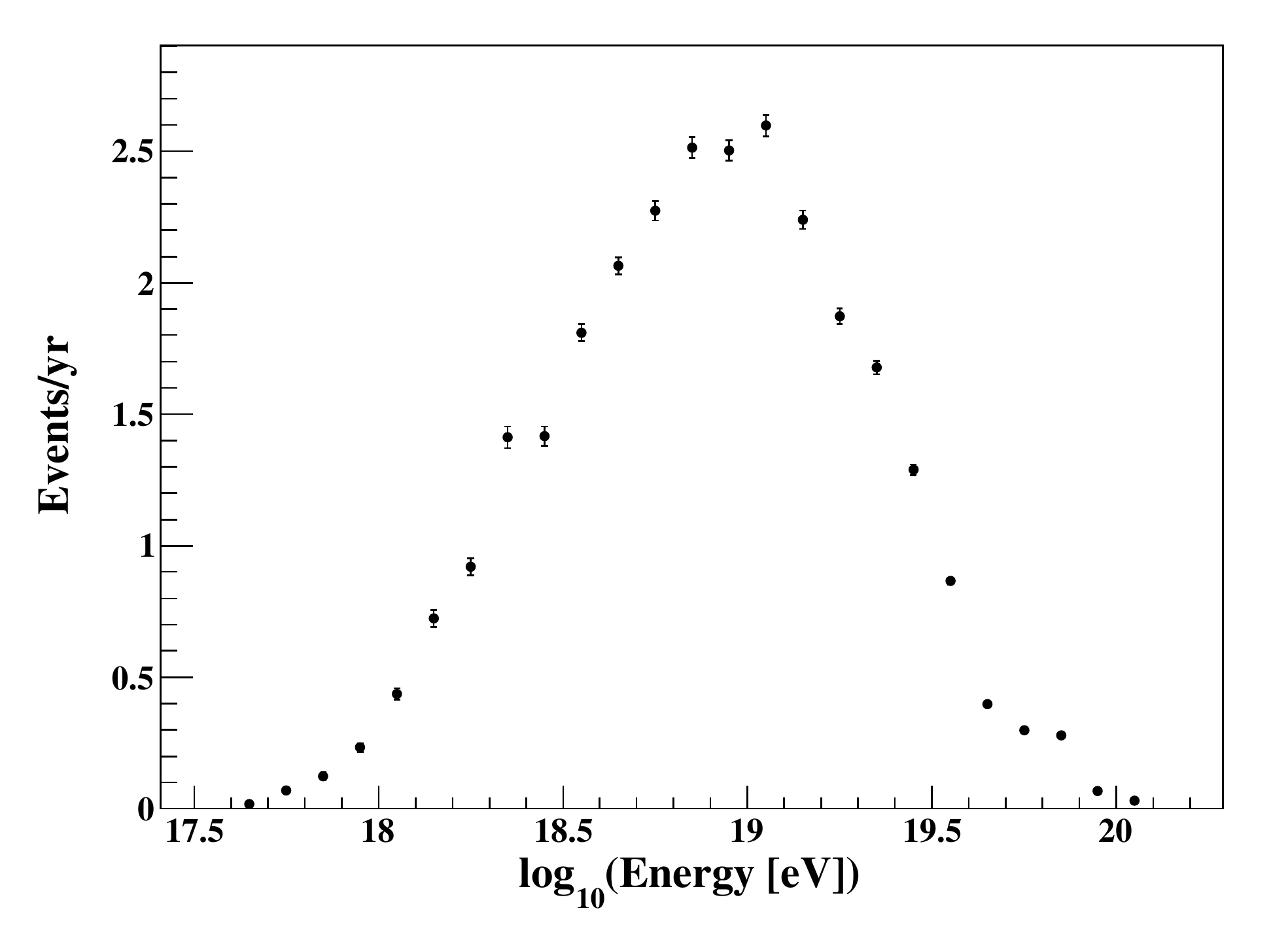}}
\caption{Expected number of triggered events per year as a function of energy, from a realistic simulation of the MIDAS telescope. A microwave flux density $F_{\rm{ref}} = F_{\rm{ref}}^0$ and a coherence parameter $\alpha=1$ were used to parameterize the EAS microwave emission in the simulation.} 
\label{spectrum}
\end{figure}

Simulations with different assumptions on the characteristics of the microwave emission from EAS were performed. For  
 $F_{\rm{ref}} = F_{\rm{ref}}^0$ and $\alpha=2$, a rate of $\sim$450 triggered events/year is expected, which reduces to $\sim$30 events/year for $\alpha=1$. The energy spectrum of the triggered events is shown in Fig.~\ref{spectrum} for the latter case.

\section{Data taking performance}
The MIDAS telescope underwent an extensive period of commissioning
during several months in 2011 at the University of Chicago, which
provided a validation of the overall design and a test of the
performance and duty cycle of the detector in a particularly
challenging environment for RF interference.  As a matter of fact,
trigger rates were found to be significantly higher than those
expected from random fluctuations. The SLT rate due to accidental
triggers, $\rm{r_{bkg}}$, is estimated to be 0.3~mHz:
\begin{equation}
\rm{r_{bkg}}=\rm{N_{patt}} \cdot \rm{n_{pix}} \cdot \left(\rm{r_{FLT}}\right)^{\rm{n_{pix}}} \left(\tau\right)^{\rm{n_{pix}}-1},
\label{eq:bkg_rate}
\end{equation}
where $\rm{N_{patt}}=767$ is the number of SLT patterns, $\rm{n_{pix}}=53$ is the number of pixels in the MIDAS camera,  $\rm{r_{FLT}}=100$~Hz is the pixel FLT rate, and $\tau = 10 ~\mu \rm{s}$ is the coincidence time window.

The background rate of SLT events during data taking was well above the estimate of Eq.~\ref{eq:bkg_rate} and highly variable, ranging from 0.01 Hz to 2 kHz. 

The major source of background was found to originate from airplanes passing over the antenna on their way to a close-by airport.  Radar altimeters on board of these aircrafts operate just above the C-Band frequency, and, while   suppressed by the MIDAS receiver bandwidth, their emissions are strong enough to produce a sudden rise of the RF background in many neighboring channels (or even in the whole camera), with a corresponding increase of the
  first and second level trigger rates over several tens of seconds. 
    
Most of the remaining background was characterized by very fast  (\textless 200 ns) transient RF signals, which trigger simultaneously from a few pixels up to the entire camera. The frequency of these events varied greatly during the data taking period, with no particular correlation with day/night or week/weekend cycles.  While their exact origin was not clear, they were likely to have an anthropogenic origin in the vicinity of the detector.  

\begin{figure}[t]
\centering{
\includegraphics[width=\columnwidth]{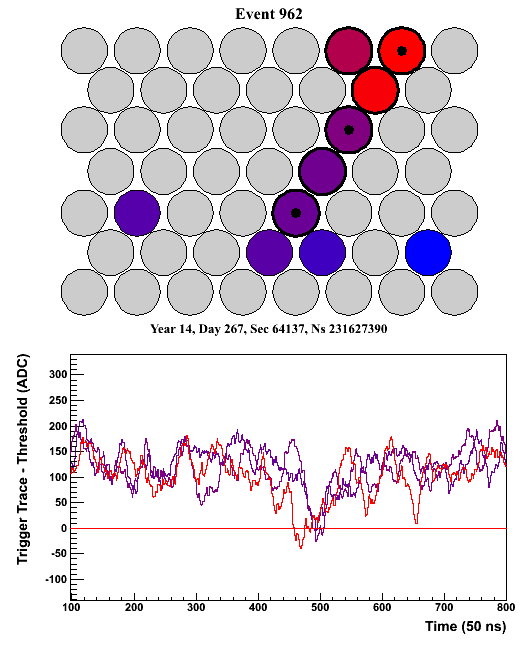}}
\caption{Event display of a candidate detected by the MIDAS telescope. See caption of Fig.~\ref{sim_event} for a description of each panel.  Although the timing characteristics of the event are compatible with those of a cosmic ray shower, correlated signals in off-track pixels do not allow for unambiguous identification.}
\label{candidate1}
\end{figure}

In the early stage of commissioning, a simple veto, which inhibited the trigger system whenever the SLT rate exceeded a given value, was implemented in the FADC firmware to avoid flooding the DAQ with background events.
Under this running conditions, the average live time of the detector was 60\%. A significant improvement was achieved with the installation of a bandpass filter designed to reject the radar frequency band, which increased the effective duty cycle of the MIDAS telescope to better than 95\%.

\begin{figure}[t]
\centering{
\includegraphics[width=\columnwidth]{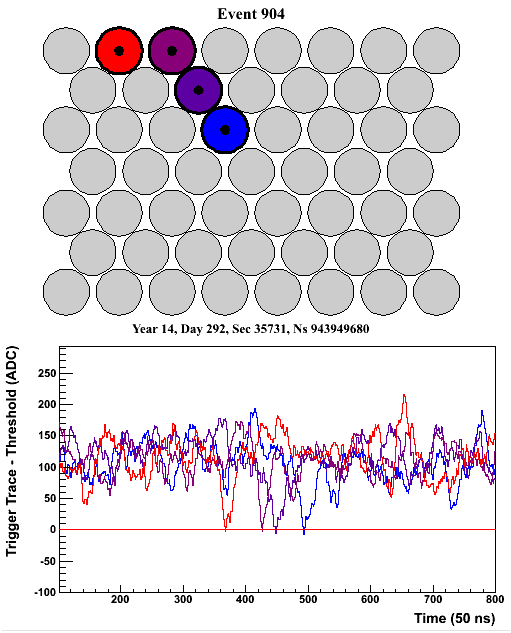}}
\caption{Event display of a 4-pixel candidate detected by the MIDAS telescope. See caption of Fig.~\ref{sim_event} for a description of each panel. Several candidates with similar characteristics - small signals and short tracks - were found in the data. }
\label{candidate2}
\end{figure}

The MIDAS telescope experienced very different weather conditions in Chicago - from high temperatures during summer, to heavy rain and storms, to snow and ice. Detector operation and data taking, as well as the telescope sensitivity, were found to be remarkably stable during the entire period of commissioning.

The main objectives of the MIDAS telescope at the University of Chicago - to validate the telescope design, and to demonstrate of a large detector duty cycle  - were successfully accomplished. In addition, a first search for events compatible with EAS was performed. Examples of candidates are shown in Figs.~\ref{candidate1}~and~\ref{candidate2}.  
However, the characteristics of these events, which have signals close to the trigger threshold, are also compatible with a tail of the overwhelming background noise. No strong conclusion can be drawn on the origin of these events until a coincident detection with well-established techniques - for example at the Pierre Auger Observatory - is performed.  A detailed account of the search and corresponding limits on microwave emission from EAS are given in~\cite{AlvarezMuniz:2012dx}.

\section{Summary and conclusions}
The MIDAS telescope - a prototype of an imaging detector for microwave emission from EAS - has been built and successfully operated at the University of Chicago.  
The telescope's design is based on inexpensive off-the-shelf microwave components, and a custom-made digital electronics and trigger system. The absolute calibration of the detector was established  with measurements of the Sun and other astrophysical objects as calibrated sources of microwave radiation. The sensitivity over the focal plane was determined with an EM simulation of the telescope validated by measurements of the Sun transit in the telescope's field of view. The sensitivity of the MIDAS detector to EAS has been studied with Monte Carlo simulations which include a realistic parameterization of the detector based on these calibration measurements. Several tens to several hundreds of events per year are expected for microwave flux intensities as suggested by laboratory measurements \cite{Gorham:2007af}.

Several months of data taking in Chicago demonstrated that MIDAS can be reliably operated with minimal maintenance, and reach a duty cycle close to 100\% even in an environment with high levels of RF background noise. Performances are expected to further improve in the radio quiet environment of the Pierre Auger Observatory, where MIDAS measurements in coincidence with the FD and SD will be essential to demonstrate the potential of this novel detection technique of UHECRs.

\section*{Acknowledgments}
This work was supported in part by the Kavli Institute for Cosmological Physics at the University of Chicago through grants NSF PHY-0114422 and NSF PHY-0551142 and an endowment from the Kavli Foundation and its founder Fred Kavli;  by the Conselho Nacional de Desenvolvimento Cient�fico e Tecnol�gico (CNPq), Brasil;  by Xunta de Galicia (INCITE09 206 336 PR), Ministerio de Educaci\'on, Cultura y Deporte (FPA 2010�18410), ASPERA (PRI-PIMASP-2011-1154) and Consolider CPAN � Ingenio2010, Spain.  The simulations used in this work have been performed on the Joint Fermilab - KICP Supercomputing Cluster, supported by grants from Fermilab, Kavli Institute for Cosmological Physics, and the University of Chicago.

\end{document}